\documentclass[range]{ar2e}

\newcommand{\gev}{\ensuremath{\mathrm{Ge\kern -0.1em V}}}
\newcommand{\gevcc}{\ensuremath{\mathrm{Ge\kern -0.1em V}/c^{2}}}
\newcommand{\ttbar}{\ensuremath{t\overline{t}}}

\newcommand{\mt}{\ensuremath{m_t}}

\usepackage{comment}

\usepackage{graphicx, subfigure}
\usepackage{amssymb}
\usepackage{epstopdf}
\begin{document}

\jname{Annual Review of Nuclear and Particle Science}
\jyear{2014}
\jvol{64}
\ARinfo{}

\title{Properties of the Top Quark}

\markboth{TopQuark}{Top Quark}

\author{Fr\'ed\'eric D\'eliot~\affiliation{CEA-Saclay, DSM/Irfu/SPP, Gif sur Yvette, France}
Nicholas Hadley~\affiliation{The University of Maryland, College Park, MD, USA}
Stephen Parke~\affiliation{Fermi National Accelerator Laboratory, Batavia, IL, USA}
Tom Schwarz~\affiliation{The University of Michigan, Ann Arbor, MI, USA}}


\begin{abstract}
The top quark is the heaviest known elementary particle, and it is often seen as a 
window to search for new physics processes in particle physics. 
A large program to study the top-quark properties has been performed both at the Tevatron 
and LHC colliders by the D0, CDF, ATLAS and CMS experiments. 
The most recent results are discussed in this review.
\end{abstract}

\maketitle


\section{Introduction}
\label{sec:introduction}

The top quark was discovered in 1995~\cite{CDF_Discovery_1995}~\cite{D0_Discovery_1995}.  It 
is the heaviest known particle, with a mass nearly equal to a tungsten atom.  Despite the incredible energy required to produce these heavy particles, measurements of top quark properties have been consistent with the Standard-Model theory
of particle physics (SM).   However, the top quark often receives special attention in new physics models because its mass requires near unity coupling to the Higgs boson.  This property is very unique among particles; therefore, studying the properties of the top quark is an important enterprise in both  measuring the SM and searching for new physics.  In this article, we will review the properties of the top quark including its mass and its charge as well as the properties of top quark decays and production.

The properties of the top quark have been studied by the CDF and D0 experiments at the Tevatron Collider at
Fermilab, and by the ATLAS and CMS experiments at the Large Hadron Collider (LHC) at CERN. 
The Tevatron Collider was a proton-antiproton collider which operated at center of mass energies of
1.8~TeV and 1.96~TeV. The LHC is a proton-proton collider which has operated at
center of mass energies of 7~TeV and 8~TeV. The CDF, D0, ATLAS, and CMS detectors all
measure charged particle momenta with central magnetic fields and charged particle tracking detectors.
All four experiments also have vertex detectors to aid in the identification of bottom quarks as
well as electromagnetic and hadronic calorimeters for measuring electrons, photons, and jets.
All the calorimeters are sufficiently hermetic so that missing transverse energy in an event can be determined.
The detectors also have extensive muon detection systems located outside the calorimeters.
Detailed descriptions of the four detectors can be found elsewhere.
~\cite{CDF_detector}~\cite{D0_detector}~\cite{ATLAS_detector}~\cite{CMS_detector}.

At both the Tevatron and the LHC, top quarks are produced primarily by the strong force in top quark-antitop quark ($t\bar{t}$) pairs.
At the Tevatron, quark-antiquark scattering dominates the top-antitop quark ($t\bar{t}$) production cross section,
while at the LHC, with its higher center of mass energy and no valence antiquarks, 
gluon-gluon scattering dominates.
The $t\bar{t}$ signatures are classified according to the decays of the W bosons that come from $t\bar{t}$ decay.
If both W bosons decay hadronically, the channel is called the alljets final state. 
The $\ell$+jets channel occurs when one W boson decays leptonically (into a muon or an electron) and the other one hadronically.
Finally when both W bosons decay leptonically, it is called the $\ell \ell$ channel.


\section{Top-Quark Intrinsic Properties}
\label{sec:intrinsic}


\subsection{Mass}
\label{mass}

The mass of the top quark (\mt) is a key parameter of the SM.
It is not predicted by the model and its large value
is affects many of the observables in top quark production and decay.
The large \mt\ is also responsible for large contributions
to quantum loop corrections to electroweak observables.
For example, together with the mass of the $W$ boson, \mt\ predicts the mass of the Higgs boson in the SM. This indirect determination can 
now be compared with the direct measurement of the Higgs boson mass to test 
the consistency of the SM.
In addition the mass of the top quark and the Higgs boson are the two parameters that
govern the shape of the Higgs potential at high energy, allowing to answer the 
fundamental question of the vacuum stability of our universe.
For all these reasons experimentally determining \mt\ as precisely as 
possible is highly important.

Since the discovery of the top quark, measurements of \mt\ have continually improved.
First determined at the Tevatron, 
\mt\ is now being measured both at the Tevatron and at the LHC with 
a precision of around 1~GeV. 
Due to this precision, the top quark has the best-known mass of all the quarks.
This precision was achieved largely 
through innovative analysis techniques including
the introduction of the so-called {\it in-situ} jet calibration.
This method is only applicable to decay channels where at least one of the $W$ bosons from the top quark 
decays hadronically.  In this technique, one calibrates the jet energy corrections by constraining the invariant mass
of the two jets from the $W$ boson to the world average measured value of the $W$ boson mass. 
Doing so significantly reduces the associated systematic uncertainty.

There are three main ways to measure \mt: (a) the template method, (b) the matrix element (ME) method, 
and (c) the ideogram method. 
Recently alternative methods have also become important (see below).

The simplest method to measure \mt\ is the template method, which relies on comparing the chosen variable that is sensitive 
to \mt\ in data with the Monte Carlo (MC) distributions (templates) generated for different values of \mt.
This observable is often the reconstructed \mt\ itself.
A maximum likelihood fit is then performed to determine the \mt\ value that best describes 
the data distribution.
In the $\ell$+jets and alljets channels, the reconstructed \mt\ can be performed using a kinematic fit to the 
\ttbar\ candidate events taking into account the different jet permutations in the \ttbar\ hypotheses. 
The maximum likelihood fit is extended to two dimensions to use the hadronically decaying $W$ boson to calibrate
the jet energy corrections. 
In the $\ell \ell$ channel, due to the presence of two undetected neutrinos, the \ttbar\ kinematics is under constrained.
Therefore,
it is necessary to make additional kinematic assumptions to be able to reconstruct \mt. 
With these assumptions, the event kinematics can be
solved and a weight for a given choice of \mt\ can be 
determined.
One can assign such weights by comparing the calculated missing transverse energy to the measured value for each event
for a given neutrino assumption (neutrino weighting technique) 
or the method can be based on the probability density of observing the measured charged lepton in the rest 
frame of the top quark (matrix weighting technique).
The statistical power of the template method can be lower compared with the other methods because it neither uses the full event information nor gives higher weights to the best-measured events.

\begin{figure}[htbp]
\begin{center}
\subfigure[]{\includegraphics[scale=0.55]{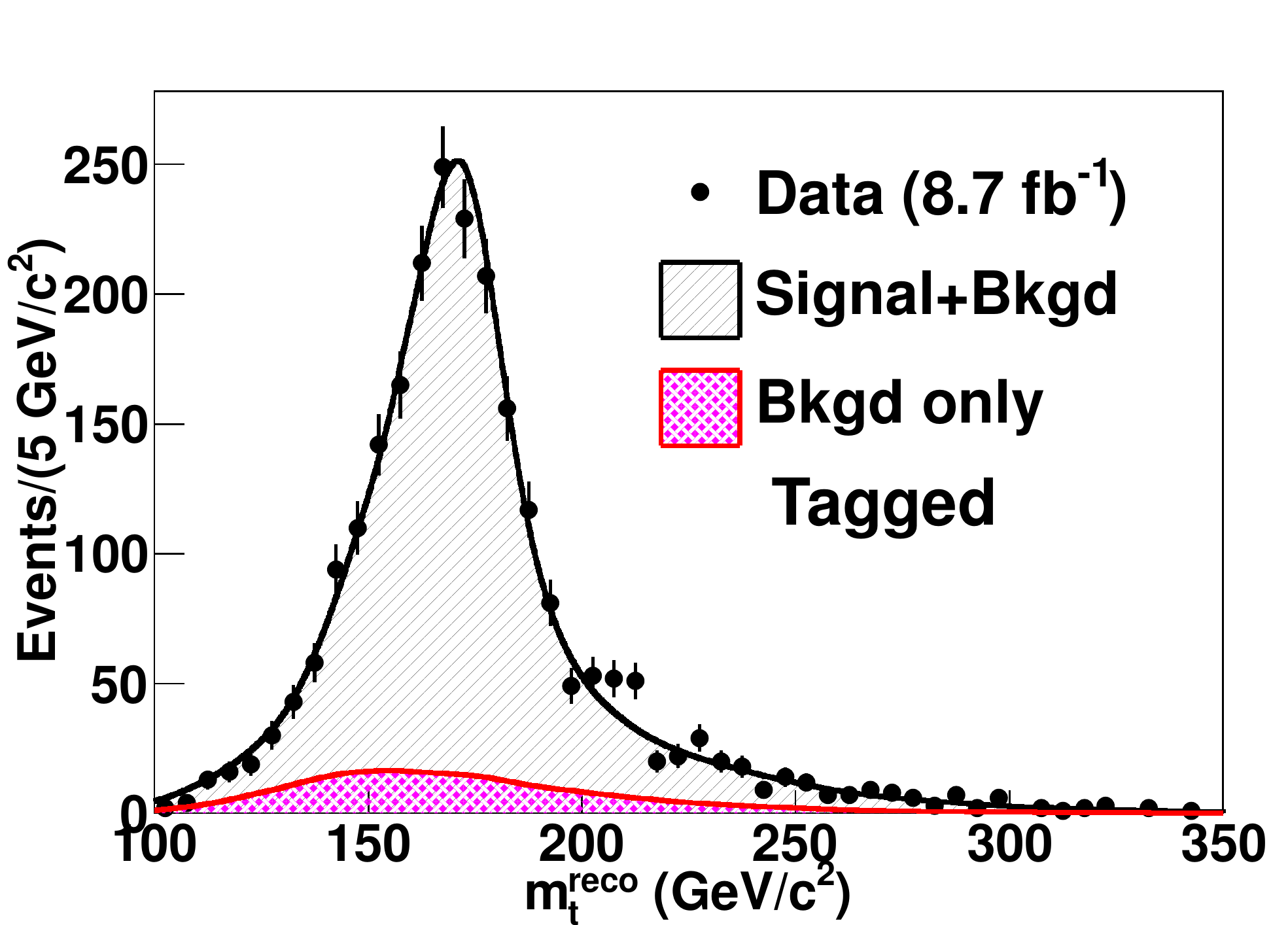}}
\subfigure[]{\includegraphics[scale=0.50]{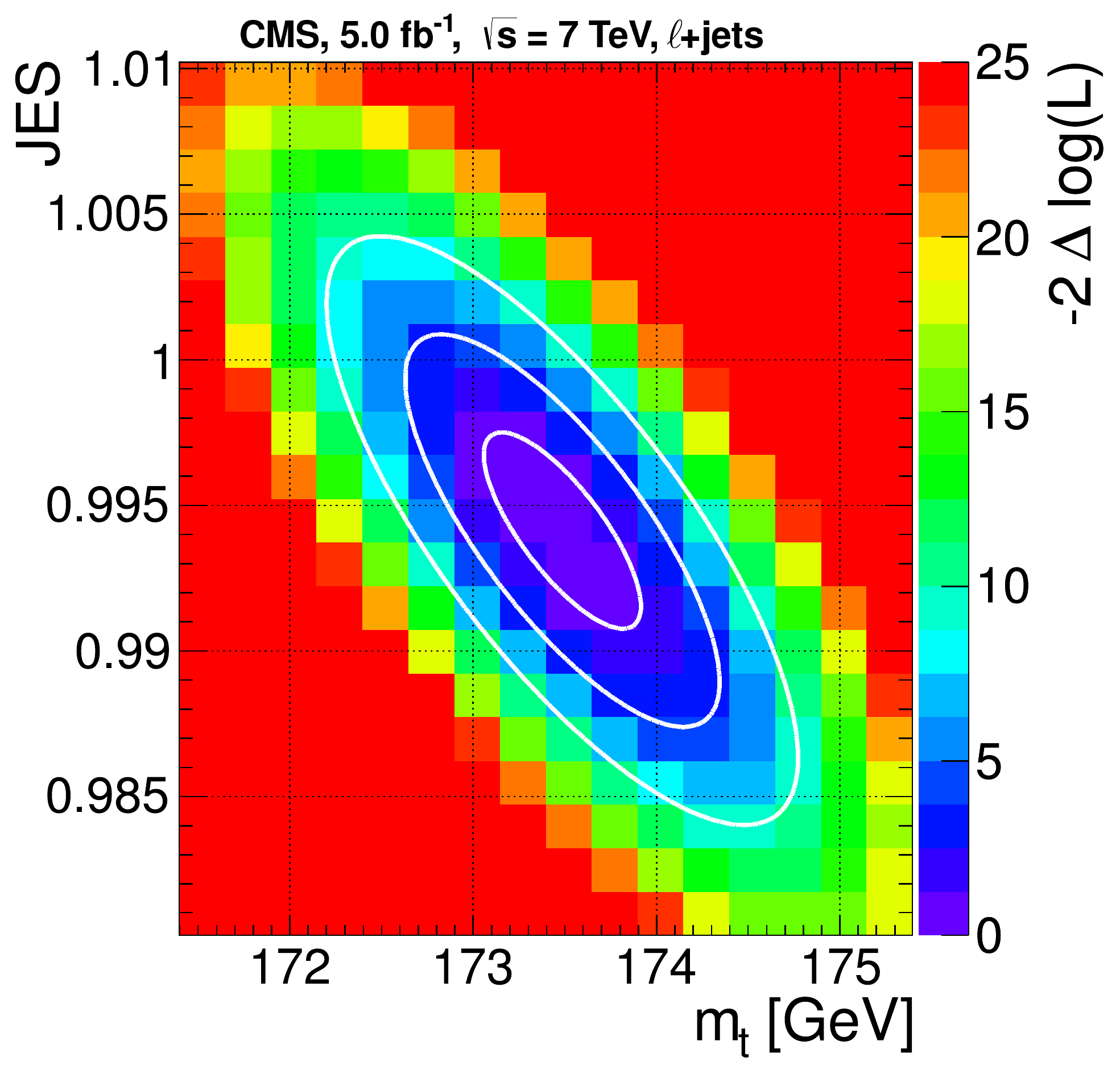}}
 \caption{(a) Distribution of the best reconstructed \mt\ in $\ell$+jets channel by CDF using a template 
method~\cite{CDF2012masslj}; 
(b) Two-dimensional likelihood to measure \mt\ with the jet energy correction in the $\ell$+jets channel 
by CMS using a ideogram method~\cite{CMS2012masslj}.}
 \label{fig:mass}
\end{center}
\end{figure}

The template method has been used to measure \mt\ in CDF, D0, ATLAS and CMS.
Table~\ref{tab:mass} summarizes the latest results for numbers that are not quoted in the text.
Figure~\ref{fig:mass} shows some examples of the distributions used to extract the results.
The latest CDF measurement in the $\ell$+jets~\cite{CDF2012masslj} uses three
observables: the best and second-best reconstructed \mt\ values 
and the invariant mass of the two jets from the hadronically 
decaying $W$ boson. The probability density functions of signal and background are estimated using a kernel density estimation method.
A template method was also employed simultaneously in the $\ell$+jets and $\ell \ell$
channels~\cite{CDF2011massdil} using
the reconstructed \mt\ from the neutrino weighting technique and a variable related to the transverse mass 
in events with two missing particles as observables for the $\ell \ell$ channel.

CDF also analyses semileptonic decaying top or antitop quarks without detection of an electron or a muon 
but with significant missing transverse energy and multiple jets~\cite{CDF2013massmet}.
The reconstruction algorithm then assumes that all selected events are $\ell$+jets \ttbar\ events with a missing particle,
the $W$ boson. The same three observables as in the resolved $\ell$+jets case 
yield $\mt = 173.93 \pm 1.64 {\rm (stat)} \pm 0.87 {\rm (syst)}$~GeV.
CDF also applies the template method in the alljets channel~\cite{CDF2012massalljets}.
D0 uses the template method in the $\ell \ell$ channel with the neutrino weighting technique~\cite{D02012massdil}.
In this analysis the jet energy calibration from the $W$ boson mass in the $\ell$+jets channel is transferred
to the $\ell \ell$ event topology.
ATLAS uses the template method in the $\ell$+jets channel in two dimensions to simultaneously determine
\mt\ with a jet energy correction factor~\cite{ATLAS2012masslj}. 
In the $\ell \ell$ channel, CMS utilizes the top-quark mass reconstructed with an analytical matrix weighting 
technique~\cite{CMS2012massdil}. This analysis uses the information provided by $b$-tagging to improve
the fraction of correctly assigned jets.

The second method, the matrix element (ME) method, is a more sophisticated technique using all measured kinematic 
quantities in the event to
construct an event-by-event probability and using the leading order (LO) matrix element integrated over the unmeasured quantities.
One takes detector effects into account by integrating resolution functions often called transfer functions.
The event probability is built from signal and background probabilities weighted by their relative contributions.
The signal probability is constructed from the convolution of the differential cross section with the parton distribution 
functions and the transfer functions. The background probability is also built using the appropriate matrix element.
With this method it is also possible to calibrate the jet energy correction {\it in-situ}.
Because it uses the full kinematic information, the ME method offers the best statistical sensitivity and was used 
at the Tevatron experiments, which the \ttbar\ statistics is less abundant than at the LHC.
However, this method is CPU intensive.
Both CDF and D0 use this technique in the $\ell$+jets and $\ell \ell$ channel. In the latest D0 analysis in the 
$\ell$+jets channel, the ME method was applied with an {\it in-situ} jet energy calibration~\cite{Abazov:2014dpa}.   
A flavor-dependent jet response correction was further applied for MC events.

The third standard method used to measure \mt\ is the ideogram method which can be considered an approximation of the 
matrix element method. Her one calculates a per-event probability of observing the reconstructed \mt\ knowing the 
resolution of this reconstructed mass and assuming a true \mt\ value.
As in the ME method, this probability is built from a signal and a background probability. 
The signal probability is obtained from a convolution of a Gaussian distribution for the mass resolution with a Breit-Wigner distribution
characterizing the decay of the top quarks, whereas the background probability is taken from MC simulation.
Typically the performance of this method falls between those of the template and the matrix element methods for limited 
\ttbar\ statistics samples and is less CPU intenssive than the ME method.
This technique was explored at the Tevatron and was most recently employed by CMS in the $\ell$+jets and alljets 
channels~\cite{CMS2012masslj,CMS2013massalljets}.
These analyses used a kinematic fit of the decay products to a \ttbar\ hypothesis and two-dimensional likelihood 
functions for each event to simultaneously estimate both the top-quark mass and the jet energy correction.
The background probability was not explicitly included in the probability expression in the $\ell$+jets channel
because the impact of the background is negligible after the final selection~\cite{CMS2012masslj}.
In the all-jets channel, this probability was estimated using an event mixing technique 
after $b$-tagging selection where jets are mixed between the different events according
to their order in transverse momentum  ($p_T$)~\cite{CMS2013massalljets}.

To further increase our knowledge of \mt, one can combine these different measurements. 
Such a combination has been performed using the BLUE method~\cite{blue1,blue2} which accounts for
the systematic uncertainty correlations of the input measurements. At the Tevatron, the latest combination, using up to 8.7~fb$^{-1}$, yields~\cite{TEV2013mass}
$\mt=173.20 \pm 0.51 {\rm (stat)} \pm 0.71 {\rm (syst)}$~GeV. This result corresponds to a total
uncertainty of 0.87~GeV and to a relative precision of 0.50\%.  
At the LHC, the latest combination, using up to 4.9~fb$^{-1}$ of data at $\sqrt{s}=7$~TeV, yields $\mt=173.29 \pm 0.23 {\rm (stat)} \pm 0.92$~GeV~\cite{LHC2013mass}. This result corresponds to a total uncertainty of 0.95~GeV and to a relative
precision of 0.54\%. The first combination of both Tevatron and LHC measurements leads to a further decrease of the total uncertainty to: $\mt = 173.3 \pm 0.27 {\rm stat)} \pm 0.71 {\rm syst}$ GeV~\cite{ATLAS:2014wva}, corresponding to a total uncertainty of 0.76 GeV and a relative precision of 0.44\%.

Despite the precision that has been achieved by the standard methods, some questions remain. Indeed, because the top quark is a 
colored object, it is difficult to know which mass is really measured. 
In all standard methods, MC is used to calibrate the measurements. The mass implemented in MC generators is different 
from a well-defined mass. A possible way out is to determine \mt\ with alternative methods. Such methods can use fewer inputs from MC or can have different sensitivity to 
systematic uncertainties compared with the standard analyses. 

One idea involves extracting \mt\ from the \ttbar\ cross section by comparing the experimental measured \ttbar\ 
cross section with the one computed theoretically. Each depends differently on \mt. The advantage of this method is that
it allows one to extract \mt\ in a well-defined renormalization scheme. However this approach is less precise
than direct measurements. D0 determined \mt\ by using the \ttbar\ cross section measured in the
$\ell$+jets channel with 5.4~fb$^{-1}$ of data; their result was $\mt=167.5^{+5.4}_{-4.9}$~GeV~\cite{D02011massfromxs}.
CMS used the measured $\ell \ell$ \ttbar\ cross section with 2.3~fb$^{-1}$ at $\sqrt{s}=7$~TeV, 
yielding $\mt=176.7^{+3.8}_{-3.4}$~GeV~\cite{CMS2013massfromxs}.

An alternative method, employed for the first time by CMS in the $\ell \ell$ channel, is called the endpoint 
method~\cite{CMS2013endpoints}. This method relies on the end distribution of the variable $m_{T2}$ used as mass 
estimator that generalizes the usual transverse mass in the case of pair produced particles with two cascade decays each 
ending in an invisible particle.
Here one extracts \mt\ by using a maximum-likelihood fit of the endpoints of three chosen $m_{T2}$ constructions, taking the 
object resolution into account.
With this technique, CMS  has measured~\cite{CMS2013endpoints}: $\mt= 173.9 \pm 0.9 {\rm (stat)} ^{+1.7}_{-2.1} {\rm (syst)}$~GeV,  
using 5.0~fb$^{-1}$. The precision of this value is already comparable to the standard measurements in the same channel.
Finally, one can also measure \mt\ by using different observables, such as the lifetime and decay length of the B-hadrons 
from the top-quark decay. The lepton $p_T$ from the decay of the $W$-boson from the 
top quark can also be used as a mass estimator.
The advantage of such estimators is that they rely only minimally on calorimeter-based uncertainties, such as the jet energy scale 
uncertainty. However, these methods may be sensitive to the modeling of
the top quark production kinematics or to the calibration of the bottom quaerk decay length or the bottom quark fragmentation model.
CDF employed such alternatives in the $\ell$+jets channel~\cite{CDF2010blifetime}, leading to precision of the order
of 4\%. 

\begin{table}
\begin{center}
\def~{\hphantom{0}}
\caption{Best top-quark mass measurement per analysis channel by the CDF, D0, ATLAS and CMS experiments.}
\label{tab:mass}
\begin{tabular}{@{}|l|l|l|@{}}
\toprule
\hline
Channel & \mt\ (GeV) & method, luminosity and Ref. \\ \hline
$\ell$+jets & CDF: $172.85 \pm 0.71 {\rm (stat)} \pm 0.85 {\rm (syst)}$ & template, 8.7~fb$^{-1}$ \cite{CDF2012masslj} \\
            & D0:  $174.94 \pm 0.83 {\rm (stat)} \pm 1.24 {\rm (syst)}$ & matrix element, 3.6~fb$^{-1}$ \cite{Abazov:2014dpa} \\ 
	    & ATLAS: $174.5 \pm 0.6 {\rm (stat)} \pm 2.3 {\rm (syst)}$  & template, 1.0~fb$^{-1}$ \cite{ATLAS2012masslj} \\
	    & CMS: $173.49 \pm 0.27 {\rm (stat)} \pm 1.03 {\rm (syst)}$ & ideogram, 5.0~fb$^{-1} \cite{CMS2012masslj} $\\ \hline
$\ell \ell$ & CDF: $170.3 \pm 2.0 {\rm (stat)} \pm 3.1 {\rm (syst)}$    & template, 5.6~fb$^{-1}$ \cite{CDF2011massdil} \\
	    & D0:  $174.0 \pm 2.4 {\rm (stat)} \pm 1.4 {\rm (syst)}$    & template, 5.3~fb$^{-1}$ \cite{D02012massdil} \\
	    & CMS: $172.5 \pm 0.4 {\rm (stat)} \pm 1.5 {\rm (syst)}$    & template, 5.0~fb$^{-1}$ \cite{CMS2012massdil} \\ \hline
alljets     & CDF: $172.5 \pm 1.4 {\rm (stat)} \pm 1.5 {\rm (syst)}$    & template, 5.8~fb$^{-1}$ \cite{CDF2012massalljets} \\ 
	    & CMS: $173.49 \pm 0.69 {\rm (stat)} \pm 1.21 {\rm (syst)}$ & ideogram, 3.5~fb$^{-1}$ \cite{CMS2013massalljets} \\ \hline
\hline
\end{tabular}
\end{center}
\end{table}


\subsection{Charge}
\label{charge}

In the SM, the top quark is the charge +2/3 isospin partner of the b quark. However,
exotic models have been proposed in which the top quark can have charge -4/3 and still decay to a W boson
and a b quark, for an example, see \cite{Chang99}.  
It remains important to determine the top quark charge experimentally
to verify that the observed top quark is the SM top quark.

The CDF~\cite{CDF2013charge} and references therein to earlier measurements, D0~\cite{Dzero2007charge}, and 
ATLAS~\cite{ATLAS2013charge} experiments have all tested the possibility that top quarks have charge -4/3
and found that the SM value is preferred. The method they used paired W bosons and bottom quarks
from top quark decays in $t\bar{t}$ events and determinee the W charge from its leptonic decay and the bottom quark charge
(whether the quark jet is from a bottom or antibottom quark) 
either from a soft-lepton tag or from the net charge of the tracks associated with the b jet. 
The current best measurement~\cite{ATLAS2013charge} has determined the top quark charge to be
$0.64\pm 0.02 {\rm (stat)} \pm 0.08 {\rm (syst)}$ and excludes charge -4/3 at more than 8~standard deviations.

\section{Top-Quark Decay Properties}
\label{sec:decay}

\subsection{Branching Ratio}
\label{sec:br}

The decay rate for $t\rightarrow Wq$ where q is a down-type quark (q= d, s, b)
is proportional to $|V_{tq}|^2$ where $V_{tq}$ is the 
Cabibbo-Kobayashi-Maskawa (CKM)~\cite{CKM}
matrix element. If we assume a unitary CKM matrix, then $|V_{tb}|$ is constrained
to be nearly equal to one, 
$|V_{tb}|$~=~$0.999152^{+0.000030}_{-0.000045}$~\cite{PDG2012}.
We define $R$ as the ratio of the branching ratios for top quark decay to Wb and top quark
decay to all types of down quarks:
$$ R\, \equiv \,  \frac{BR(t\rightarrow Wb)}{BR(t\rightarrow Wq)}\, \, = \, \,
\frac{|V_{tb}|^{2}}{ |V_{tb}|^{2} + |V_{ts}|^{2} + |V_{td}|^{2} }.
$$
Assuming three generations and given our knowledge of $V_{ts}$ and $V_{td}$,
R should also be nearly equal to one, 
R~=~$0.99830^{+0.00004}_{-0.00009}$~\cite{PDG2012}.
New physics, such as a fourth generation, could cause the measured value
for R to differ from the prediction.

Both the CDF~\cite{CDF2013BR}and references therein 
and D0~\cite{Dzero2011BR}and references therein experiments have both measured R using $t\bar{t}$ events
in the $\ell$+jets channel.
Their basic method is to measure the number of 
$t\bar{t}$
events with zero, one, or more than
one jet that is identified as a b-quark jet. These numbers can be used to 
reduce the systematic uncertainty due to the uncertainty on the b-tagging
efficiency and to simultaneously measure R and the  $t\bar{t}$ production 
cross section. The current measurement from D0
uses the combined information
from $t\bar{t}$ events in the $\ell$+jets and $\ell \ell$ channels. 
D0 has measured R~=~$0.90\pm0.04$, which is ~ 2.5 $\sigma$
from the SM prediction. CDF's~\cite{Aaltonen:2014yua} latest result in the $\ell$ channel provides $R = 0.93 \pm 0.04$, whereas CMS~\cite{Khachatryan:2014nda} has measured $R= 1.01 \pm 0.03$; both are consistent with the SM prediction.

\subsection{Width}
\label{sec:width}
Because \mt\ is well above the Wb threshold, the top quark
width is expected to be dominated by the decay $t\rightarrow Wb$.
Neglecting higher-order weak corrections and terms of $m_{b}^{2}/m_{b}^{2}$,
the SM predicts the top quark width, ($\Gamma_t$) at next-to-leading order (NLO) 
to be~\cite{Jezabek1989width}

$$\Gamma_{t} = \frac{G_{F} m_{t}^{3}}{8 \pi \sqrt{2}}\,  |V_{tb}|^{2} \,   
(1 - \frac{M_{W}^{2}}{M_{t}^{2}})^{2} \, \,
(1 + 2 \frac{M_{W}^{2}}{M_{t}^{2}})^{2} \, \,
[1 - \frac{2 \alpha_{s}}{3 \pi}(\frac{2 \pi^{2}}{3} - \frac{5}{2}) ]
$$

where $M_W$ the mass of the W boson, $m_b$ is the mass of the bottom quark, $G_F$ the Fermi coupling constant,
$\alpha_s$ is the strong coupling constant, and $V_{tb}$ the CKM ME that provides the strength of the left-handed Wtb coupling. 
Assuming $V_{tb}$~=~1 and a top-quark mass
of 172.5~GeV, the value of $\Gamma_t$ at NLO is 1.33 GeV.
A recent next-to-next-to-leading-order (NNLO) calculation 
predicts $\Gamma_t$~=~1.32~GeV~\cite{Gao2013width}.
Of course, new physics can alter the value of $\Gamma_t$ from that predicted by the SM.

Both CDF and D0 have measured $\Gamma_F$.
CDF measures the
width directly using the reconstructed top-quark mass for $t\bar{t}$ events
in the $\ell$+jets channel~\cite{CDF2013width}. 
They determine the energy scale for calorimeter jets using an 
{\it in situ} calibration and an artificial neural network to improve the jet-energy calibration.
In addition to the jet-energy scale, the other dominant systematic uncertainties come from the choice
of the event generator and from uncertainties in color reconnection.
For \mt = 172.5~GeV, CDF finds, using 8.7~fb$^{-1}$ of data, $1.10 < \Gamma_{t} < 4.05$~GeV at
68\%\ CL.

D0 determines $\Gamma_{t}$ using the partial decay width $\Gamma(t\rightarrow Wb)$
and the branching ratio, $BR(t\rightarrow Wb)$~\cite{Dzero2012width}.
The single top-quark production cross
section in the $t$-channel, $\sigma$(t~-~channel),
is used to determine $\Gamma(t\rightarrow Wb)$ and
 $BR(t\rightarrow Wb)$ is measured in $t\bar{t}$ events. The analysis then uses the
 values of $\sigma$(t~-~channel) and  $\Gamma(t\rightarrow Wb)$ as calculated in the SM
 to obtain $\Gamma_{t}$.
 Using 5.4~fb$^{-1}$ of data, D0 obtains $\Gamma_{t} = 2.00^{+0.47}_{-0.43}$~GeV.
 This result is more precise than the direct measurement by CDF, but has
 more dependence on theoretical assumptions.

\subsection{W Helicity in Top-Quark Decays}
\label{sec:whelicity}
In the decay $t\rightarrow Wb$, because the W bosons are real particles, their polarizations can be
left-handed, right-handed or longitudinal. The fraction of decays with a particular polarization are referred to
as the helicity fractions and are denoted 
$F_L$, $F_R$, and $F_0$ for left-handed, right handed, and longitudinal W polarizations
respectively. 
For unpolarized top-quark production, in the SM, the helicity fractions
are approximately 70\%\ longitudinal and 30\%\ left-handed. At LO and in the limit
of zero b-quark mass, helicity suppression forces $F_R$ to vanish. Higher order corrections modify
these predictions slightly. 
Current NNLO calculations~\cite{Czarnecki2010} yield $F_{L} = 0.311 \pm 0.005$, $F_{R} = 0.0017 \pm 0.0001$, and
$F_{0} = 0.687 \pm 0.005$ for \mt = $172.8 \pm 1.3$~GeV.
The helicity fractions can be measured through studies of the angular distributions of the top quark decay
products in $t\bar{t}$ events. Th helicity angle, $\theta^*$, is defined as the angle between
the W-boson momentum in the top-quark rest frame and the
momentum of the down-type fermion from the W decay measured in the rest frame of the W boson.
The angular distribution is~\cite{Kane1992}:
$$ \frac{1}{\sigma}  \frac{{\rm d} \sigma}{{\rm d} \cos \theta^{*}} = \frac{3}{8} F_{L} (1 - \cos \theta^{*})^{2}
+ \frac{3}{8} F_{R} (1 + \cos \theta^{*})^{2}
+ \frac{3}{4} F_{0} (\sin \theta^{*})^{2}.
$$

Anomalous Wtb couplings will cause the helicity fractions and the angular distributions to deviate from their
SM values. In effective field theories, one can introduce dimension-six operators that modify
the Wtb vertex, and one can constrain the coefficients that specify the strength of these anomalous couplings by measuring the helicity 
fractions~\cite{Aguilar2009WPOL}~\cite{Zhang2011WPOL}.

The CDF, 
D0, 
ATLAS, 
and CMS experiments 
have all measured the W boson polarizations in top quark decays
in $t\bar{t}$ 
events~\cite{CDFDzero2012WPOL},~\cite{ATLAS2012WPOL},~\cite{CMS2013WPOL}.
The experiments have studied top-quark decays in the $\ell$+jets and $\ell \ell$ channels.
The constraint $F_{L} + F_{0} + F_{R} = 1 $ is used in determining the helicity fractions when
fitting the angular distributions. The measurements from all experiments agree with SM
predictions. The current best values for the helicity fractions are from CMS
based on 5.0 fb$^{-1}$ of data at 7~TeV and are:
$$F_{L} = 0.319 \pm \ 0.022\, {\rm (stat.)} \pm 0.022\, {\rm (syst.)}, $$
$$F_{0} = 0.682 \pm \ 0.030\, {\rm (stat.)} \pm 0.033\, {\rm (syst.)} , $$ 
$$F_{L} = 0.008 \pm \ 0.012\, {\rm (stat.)} \pm 0.014\, {\rm (syst.)} ,$$
with a correlation coefficient of -0.95 between $F_0$ and $F_L$~\cite{CMS2013WPOL}.

\subsection{Flavor Changing Neutral Current}
\label{sec:fcnc}
In the SM, the top quark is expected to decay nearly all of the time
to a W boson and a bottom quark. The flavor-changing neutral-current decay of the
top quark is suppressed
by the GIM mechanism in a similar manner to those of other quarks. The decay of a top quark to a Z boson and a up or charm quark
occurs only through higher-order diagrams with loops, and the branching ratio for $t\rightarrow Zq$ is
predicted
to be of order $10^{-14}$~\cite{JAAS04}. 
However, extensions to the SM can predict enhancements to the 
branching ratio (BR) for $t\rightarrow Zq$ and the CDF~\cite{CDF2008FCNC}, 
D0~\cite{Dzero2011FCNC}, 
ATLAS~\cite{ATLAS2012FCNC}, and 
CMS~\cite{Chatrchyan:2013nwa} experiments
have all searched for the decay in $t\bar{t}$ events. In these searches, one top quark is 
assumed to decay to $Wb$ and the other to $Zq$ where q can be an up or charm quark.
The Z bosons are identified by their decays to $e^{+}e^{-}$ or $\mu^{+}\mu^{-}$
pairs, and the W bosons are required to decay 
to a electron or muon and a neutrino. 
None of the experiments observed an
excess of events over standard model backgrounds. The best current limit 
is that BR($t\rightarrow Zq$) greater than 0.05\%\ is excluded at 95\%\ CL. This limit is from CMS~\cite{Chatrchyan:2013nwa} and is based on 
5.0 fb$^{-1}$ and 19.7 fb$^{-1}$ of data collected
at 7 TeV and 8 TeV, respectively.


\section{Top-Quark Properties dependent on Production}
\label{sec:production}

\subsection{Top-Quark Polarization}
\label{sec:polarization}

The top-quark decay width ($\sim$  1 GeV, see Section~\ref{sec:width}) is much larger than the QCD hadronization scale ($\Lambda_{QCD} \sim $ 0.1 GeV) and much larger than the spin decorrelation scale ($\Lambda_{QCD}^2/m_t \sim$ 0.1 MeV). Therefore, any polarization of the top-quark or any spin correlations in top quark pair production are reflected in angular correlations of the decay products~\cite{Mahlon:1995zn,Stelzer:1995gc}.

The decay products from a polarized top quark have their moment vectors correlated with the top quark spin axis as follows:
\begin{eqnarray}
\hspace*{-0.25cm} \frac{1}{\Gamma_T} \frac{d \Gamma}{d \cos \chi_i} 
= (1+\alpha_i \cos \chi_i)/2, ~~ {\rm where} ~\alpha_i=\left\{ \begin{array}{lll}
+1.0 &(+0.998)   &\hbox{$l^+$ }\\
+1.0 &(+0.966)  &\hbox{$\bar{d}$-quark}\\
-0.31 &(-0.314)  &\hbox{$\bar\nu$ } \\
-0.31 &(-0.317)   &\hbox{$u$-quark} \\
-0.41 &(-0.393)   &\hbox{$b$-quark}
  \end{array} \right.
\nonumber
\end{eqnarray}
at LO with the NLO results in parenthesis~\cite{Bernreuther:2001rq}. Here $\chi_i$ is the angle between the decay product and the spin axis in the top quark rest frame. The net polarization of the top quark can be measured from the correlations of decay products with the spin axis:
\begin{eqnarray}
\frac{1}{\sigma} \frac{d \sigma}{d \cos \chi_i} 
= (1+P_{t} ~\alpha_i \cos \chi_i)/2
\nonumber
\end{eqnarray}
where $P_{t}$ is the top quark polarization.  In the SM, this is expected to be small, $P\approx 0.003$, and driven by electroweak correction~\cite{Bernreuther:2013aga}.

ATLAS has measured the production of $P_{t} \alpha_l$ for the two possibilities that top and anti-top have the same polarization [i.e., are CP conserving (CPC)] and that they have opposite polarization [i.e., are CP violating(CPV)]~\cite{Aad:2013ksa}. The results are
\begin{eqnarray}
P_t^{CPC} \alpha_l  & = -0.035 \pm 0.014 \pm 0.037 \quad {\rm and} \quad
P_t^{CPV} \alpha_l  & = 0.020 \pm 0.016 ^{+0.013}_{-0.017}.
\nonumber
\end{eqnarray}
For CMS, the top quark polarization $P$ in the helicity basis is given by $P_t=2A_P$ where:
\begin{eqnarray}
A_P & =  &\frac{N(\cos \chi_l >0) - N(\cos \chi_l <0)}{N(\cos \chi_l >0) + N(\cos \chi_l <0)} 
\nonumber
\end{eqnarray}
$A_P$ has a measured value of  $0.005 \pm 0.013 \pm 0.020 \pm 0.008 $, assuming CP 
invariance~\cite{Chatrchyan:2013wua}.

\subsection{\ttbar\ Charge Asymmetry}
\label{sec:asymmetry}



Measuring the charge asymmetry in top-quark production is a test of discrete symmetries of the strong interaction.  In the $t\bar{t}$ center of mass frame, top quarks (antitop quarks) are produced preferentially in the direction of the incoming quark (antiquark) because of an NLO QCD effect that is present only for asymmetric initial states such as $q\bar{q}$ and $qg$.  
The predicted SM asymmetry at the Tevatron is quite modest, and even smaller at the LHC.  However, new physics, such as a new boson with a charge or parity violating component, could enhance the effect and result in a much larger measured asymmetry~\cite{asym_theory1, asym_theory2, asym_theory3, asym_theory4, asym_theory5, asym_theory6, asym_theory7}.  

\begin{figure}
  \begin{center}
  \includegraphics[width=0.8\textwidth]{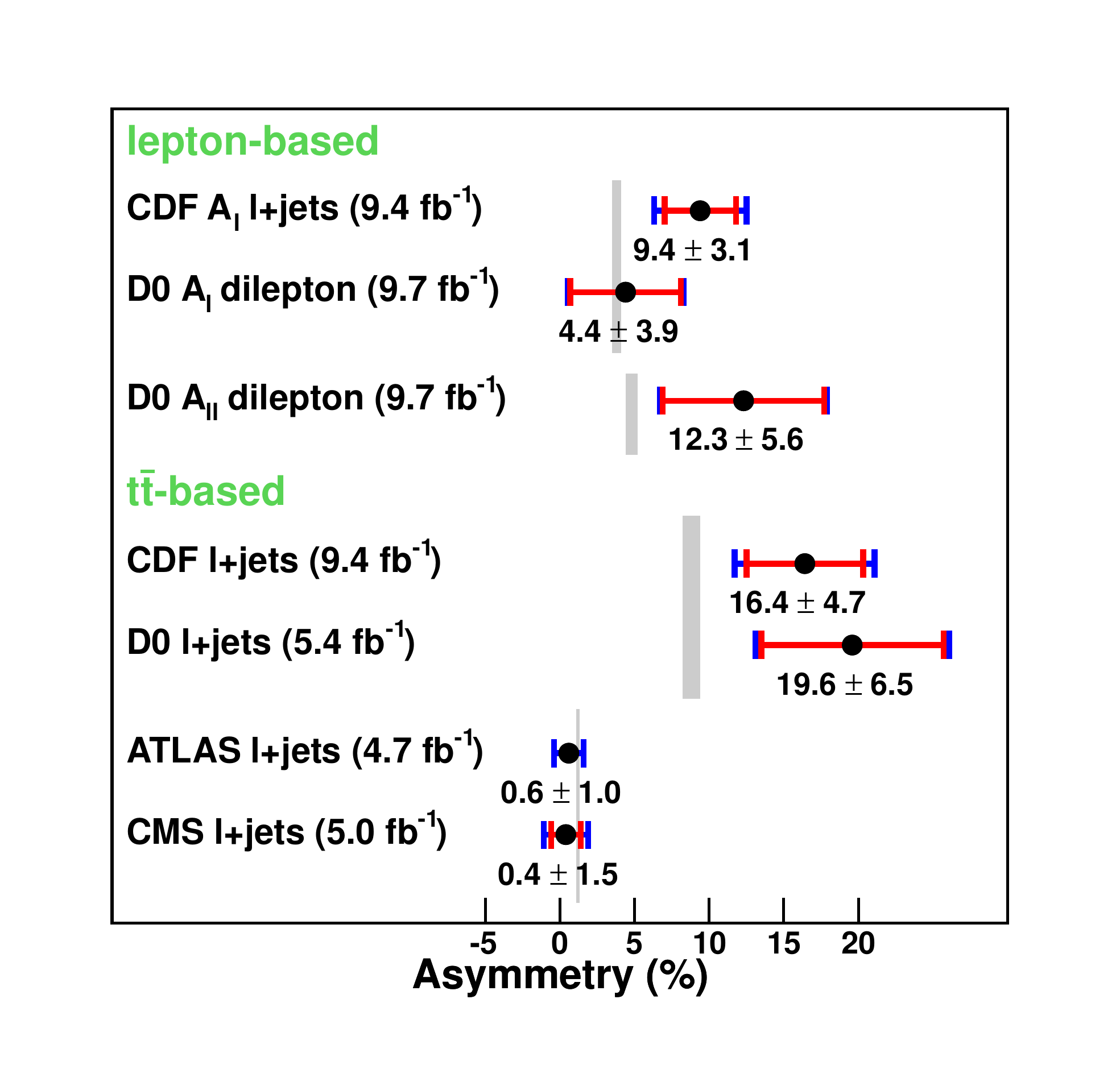}
\caption{Recent measurements of the top quark forward backward asymmetry at the Tevatron, and the forward charge asymmetry at the LHC. The grey bands show the standard model predictions. \label{ac_results}}
    \end{center}
\end{figure}

Because the top quarks at the Tevatron are produced in an asymmetric (proton-antiproton) initial state, the charge asymmetry manifests itself as a forward-backward asymmetry.

\begin{equation}
A_{FB} = \frac{N(\Delta y > 0) - N(\Delta y < 0)}{N(\Delta y > 0) + N(\Delta y < 0)}
\nonumber
\end{equation}

\noindent where $\Delta y$ is the rapidity difference between the top and the anti-top quark, and forward (backward) in the direction of the incoming proton (antiproton). The most recent calculations at NLO including electroweak corrections predict $\mathrm{A_{FB} = 8.8 \pm 0.6\%}$ at the Tevatron ($\sqrt{s} = 1.9$ TeV proton-antiproton collisions)\cite{asym_prediction1,asym_prediction2}.  The measurement of the top quark charge asymmetry has generated a great deal of interest within the past decade because measurements of the asymmetry at both the CDF and D0 experiments have been somewhat larger than the SM prediction~\cite{cdf1,cdf2,cdf3,cdf4,d01,d02,d03,d04}.  
Measurements at CDF have reported that the measured asymmetry depends on event kinematics.  In particular, the asymmetry is larger at higher $m_{t\bar{t}}$ and $|\Delta y|$.  D0 observed no significant increase at larger $m_{t\bar{t}}$. Figure \ref{ac_results} summarizes the latest inclusive measurements.

\begin{figure}[ht]
\begin{minipage}[b]{0.45\linewidth}
\centering
\includegraphics[width=0.8\textwidth]{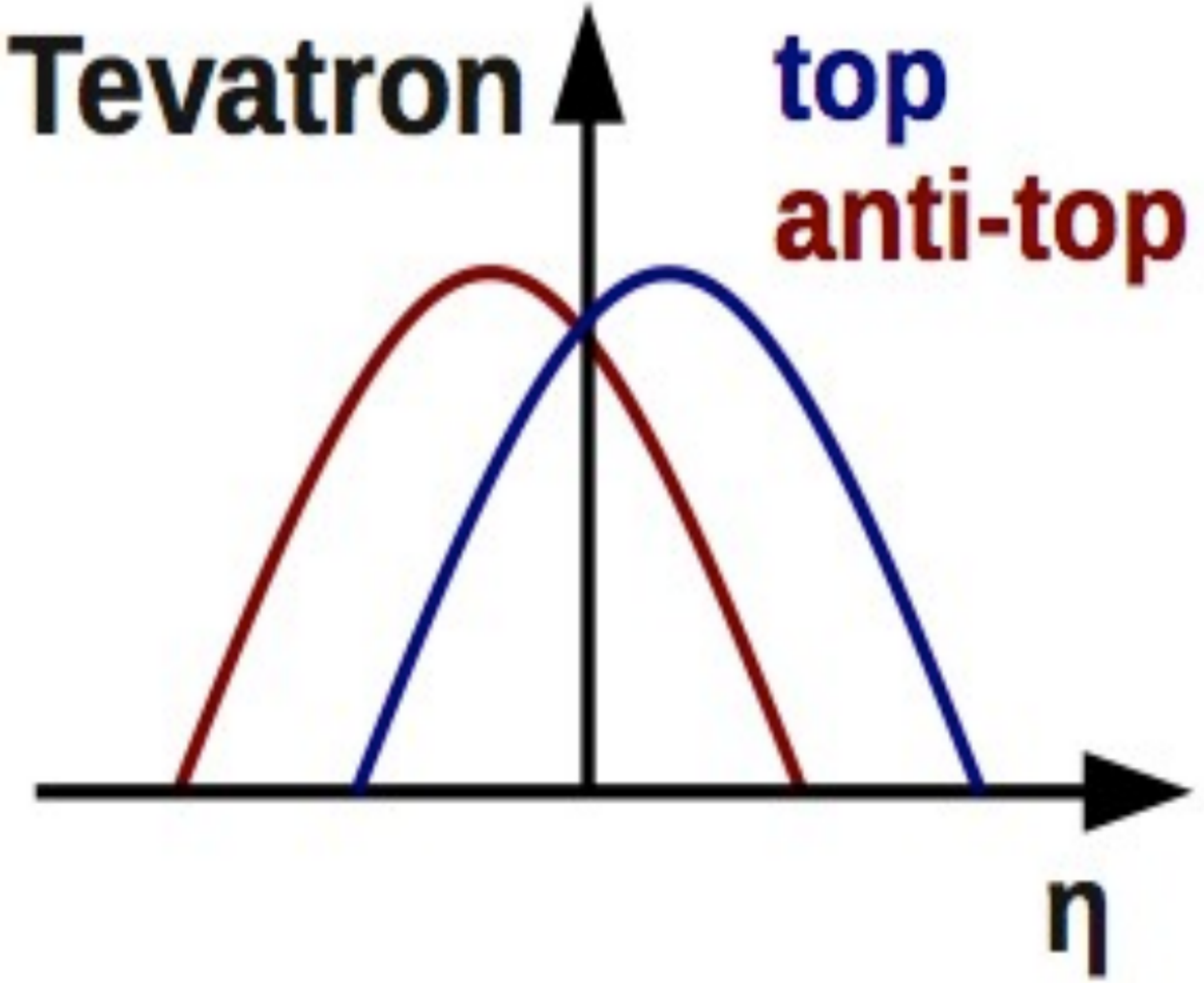}
\end{minipage}
\hspace{0.5cm}
\begin{minipage}[b]{0.45\linewidth}
\centering
\includegraphics[width=0.8\textwidth]{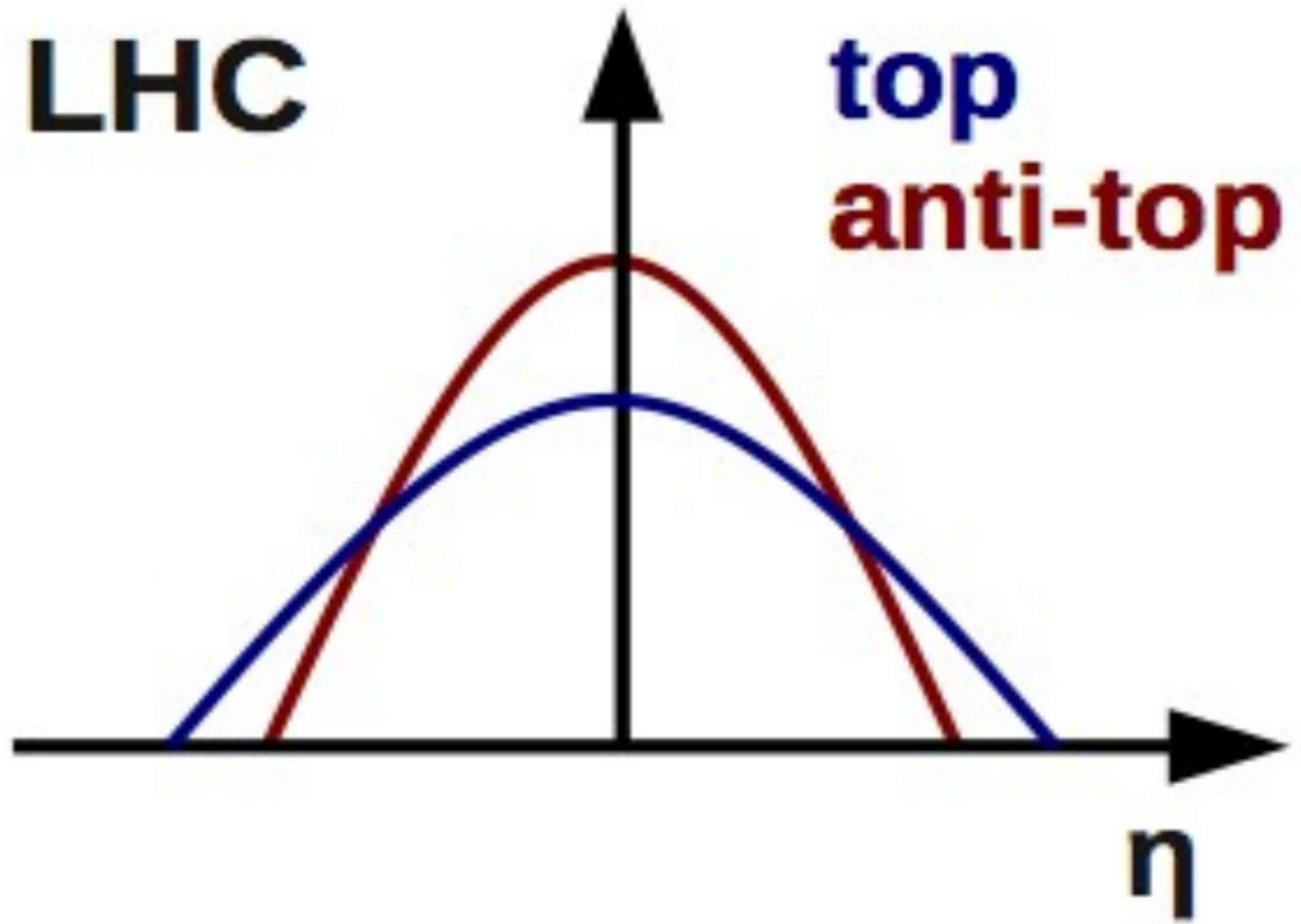}
\end{minipage}
\caption{Graphical representation of top quark and anti-top quark rapidity distributions at the LHC (p-p collisions) and the Tevatron (p-$\mathrm{\bar{p}}$ collisions). \label{ac_image}}
\end{figure}

At the LHC, the symmetric proton-proton collisions do not define a forward and backward direction.  Furthermore, top quarks are usually produced from gluon-gluon fusion, which does not lead to charge asymmetry.  However, a fraction of collisions do come from quark-antiquark interactions, in which the antiquark originates from the proton sea.  The antiquarks from the sea tend to have far less momentum than do valence quarks, causing top (antitop) quarks to be preferentially produced at higher (lower) rapidity.  Figure \ref{ac_image} depicts the distributions of top and antitop quarks as a function of rapidity for the Tevatron and LHC.

\begin{figure}
  \begin{center}
  \includegraphics[width=0.75\textwidth]{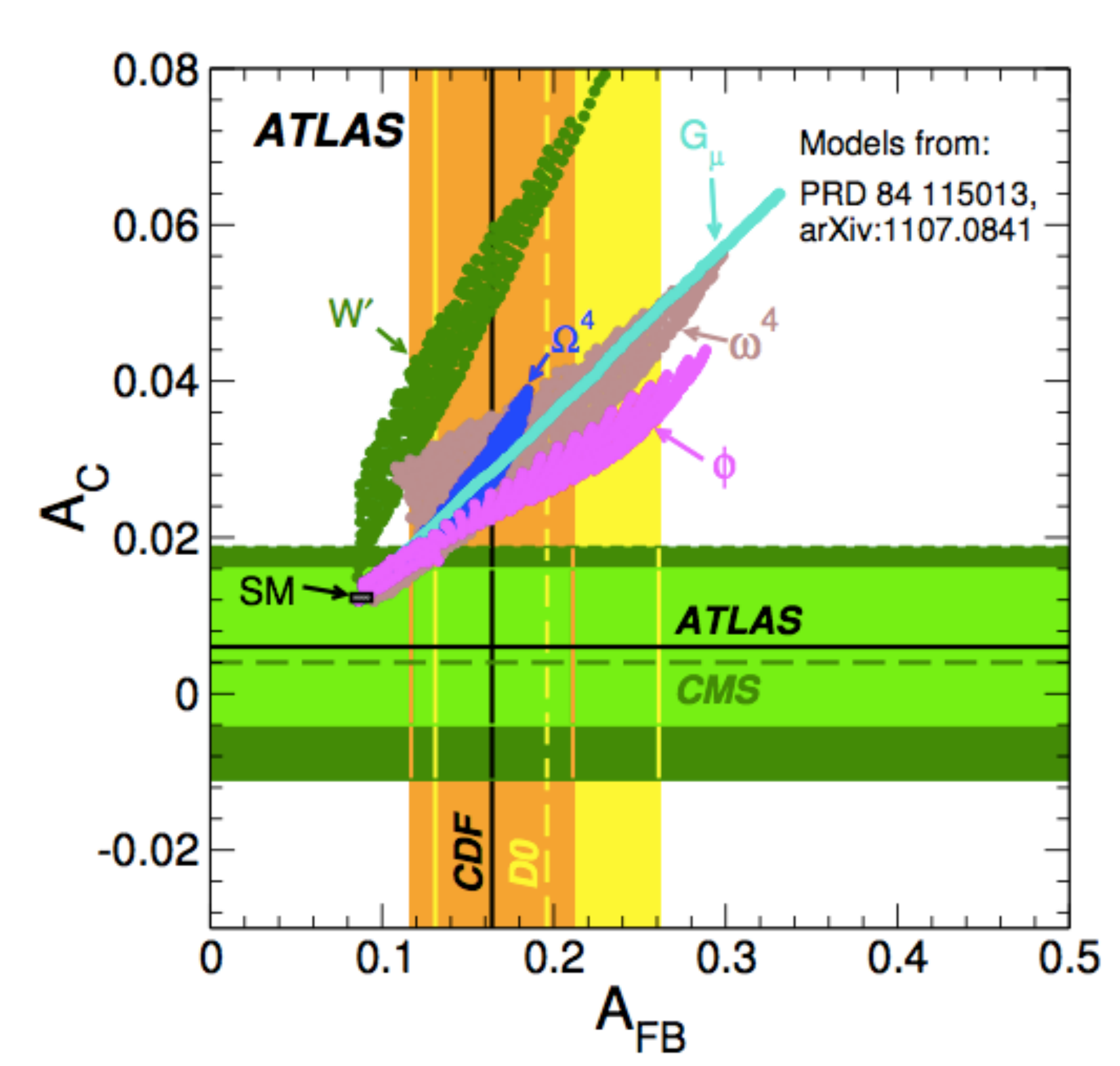}
\caption{Predictions of the charge asymmetry at the LHC and the Tevatron for several new physics models compared to current measurements \cite{saveedra}. \label{models_results}}
    \end{center}
\end{figure}

The charge asymmetry can then be probed by measuring

\begin{equation}
A_C = \frac{N(\Delta |y| > 0) - N(\Delta |y| < 0)}{N(\Delta |y| > 0) + N(\Delta |y| < 0)}
\nonumber
\end{equation}

\noindent where $\Delta |y| = |y_t| - |y_{\bar{t}}|$ is the difference between the magnitude of the top quark rapidity and the magnitude of the anti top quark rapidity.  Recent calculations at NLO, including electroweak corrections, predicted $\mathrm{A_{C} = 1.23 \pm 0.05\%}$ at the LHC ($\sqrt{s} = 7$ TeV for pp collisions) \cite{asym_prediction2}.  ATLAS and CMS measurements of the charge asymmetry (at $\sqrt{s} = 7$ utilizing 5 $\mathrm{fb^{-1}}$ of data) have not observed any significant deviation from the SM (Figure \ref{models_results}~\cite{asym_cms1,asym_cms2,asym_cms3,asym_atlas1,asym_atlas2}).  
ATLAS and CMS have also performed differential measurements as a function of $m_{t\bar{t}}$, rapidity, and transverse momentum, and have not observed deviations from the SM.  However, because the data are dominated by $t\bar{t}$ events produced from gluon fusion, uncertainties on these measurements are still too large to make a definitive statement about the Tevatron results.  


Measurements at the LHC and the Tevatron are performed similarly.  All current published analyses selected $t\bar{t}$ events in the semileptonic decay channel, by requiring events with exactly one isolated, high-$p_T$ lepton; a large amount of missing transverse energy; and several (three or more) high-$p_T$ jets.  

A topological algorithm is used to reconstruct the kinematics of both the top and antitop quarks from the observed decay products.  To properly reconstruct the masses of the top quark, the antitop quark, and the intermediary decaying W bosons, this algorithm decides which jets must match to which partons.
The momentum along the beam axis is also reconstructed.  Reconstructing the top events is of course not perfect.  Approximately 50\% of events at both the Tevatron and LHC simply lose a parton from $t\bar{t}$ decay during selection.  Furthermore, in events wherein all partons were found in jet selection, the reconstruction algorithm matches them to the correct jets with an efficiency of roughly 60-70\%.  A regularized unfolding procedure is employed in all analyses to correct for the smearing of the top quark kinematics.  
To first order, the unfolding technique can be though of as matrix inversion from the reconstructed rapidity distribution to the corrected (unfolded) rapidity distribution.  The unfolding ME quantifies the bin-by-bin smearing for $t\bar{t}$ events, which is built from MC simulations.  The asymmetry is then measured simply from the unfolded $\Delta y$ distribution.  The predicted distribution of $\Delta y$ for backgrounds is subtracted from data before unfolding.  When properly employed, the unfolding technique is fairly model independent, as long as kinematic distributions that are correlated to reconstruction efficiency are similar to the MC simulation used to derive the smearing matrix.


Currently, published measurements of the charge asymmetry are still statistically limited even when using the full 5 $\mathrm{fb^{-1}}$ of 7 TeV data for both the CMS and ATLAS experiments.  However, systematic uncertainties are expected to be dominant for measurements of the charge asymmetry using the 20 $\mathrm{fb^{-1}}$ data set at 8 TeV, as well as for anticipated higher-luminosity measurements (300 $\mathrm{fb^{-1}}$ and 3,000 $\mathrm{fb^{-1}}$) with the data sets at 14 TeV.  Systematic uncertainties taken into account in measurements of the charge asymmetry include those from detector effects (such as jet energy scale/resolution, lepton energy scale/resolution, and pileup), modeling (such as MC generator and hadronization, MC-derived backgrounds, and parton distribution function uncertainties), and measurement techniques (such as unfolding).  The leading two systematics in each measurement are either simulation modeling uncertainties or calibration uncertainties for leptons and jets.  Without improvements in technique, these uncertainties will not necessarily scale with integrated luminosity.  


The pseudorapidity of charged leptons in $t\bar{t}$ event candidates can also be used to probe the charge asymmetry.  Although the predicted asymmetry is much smaller ($\sim 0.4\%$), this technique prevents complications associated with reconstructing and unfolding the top and anti-top quarks in the event.  The lepton-based asymmetry is performed similarly to the reconstructed top quark measurements.  Both CDF and D0 have measured the single-lepton asymmetry ($A_l$) in top quark events:

\begin{equation}
A_{l} = \frac{N(q_l \cdot \eta_l > 0) - N(q_l \cdot \eta_l < 0)}{N(q_l \cdot \eta_l > 0) + N(q_l \cdot \eta_l < 0)}
\nonumber
\end{equation}

\noindent where $q_l$ is the charge of the lepton and $\eta_l$ the pseudo rapidity.  D0 has also probed the charge asymmetry using both leptons in $\ell \ell$ top quark decays.  The $\ell \ell$ asymmetry is defined as

\begin{equation}
A_{ll} = \frac{N(\Delta \eta > 0) - N(\Delta \eta < 0)}{N(\Delta \eta > 0) + N(\Delta \eta < 0)}
\nonumber
\end{equation}

\noindent where $\Delta \eta = \eta_{l+} - \eta_{l-}$ is the rapidity difference between the positive and negative charged leptons. Figure \ref{ac_results} shows results from CDF~\cite{cdfAl,Aaltonen:2014eva}  and D0~\cite{d0Al,Abazov:2014oea} for the lepton-based asymmetries.  Similar to Tevatron results for the reconstructed top quark asymmetry, the lepton-based results are within two $\sigma$ of the SM prediction.  CMS has also recently published an asymmetry measurement using leptons~\cite{Chatrchyan:2014yta}, in agreement with the SM prediction.

As note above, some of the measurements by CDF and D0 show a top quark forward-backward asymmetry that is larger than predicted by the SM, and CDF has published evidence that the asymmetry is larger for higher invariant $t\bar{t}$ mass leading to a great interest in the theoretical community.
Following Reference~\cite{saveedra}, we classify the proposed new physics models possibly that may explain the results as having the following: 

\begin{itemize}
\item $Z^{\prime}$: A neutral color isospin-singlet vector boson exchanged in $u\bar{u} \rightarrow t\bar{t}$ through the t-channel.
\item $W^{\prime}$: A charged color and isospin singlet vector exchanged in $d\bar{d} \rightarrow t\bar{t}$ through the t-channel.
\item $\mathcal{G_\mu}$: A neutral color-octet boson with axial vector couplings (axigluon) exchanged in $q\bar{q} \rightarrow t\bar{t}$ through the s-channel.
\item $\phi$: A scalar doublet which contains neutral and charged scalars exchanging top quarks to the first generation through the t-channel. 
\item $\omega^4$: Color-triplet scalar with charge 4/3, containing both neutral and charged scalars.  Exchanged in $u\bar{u} \rightarrow t\bar{t}$.
\item $\Omega^4$: Color sextet scalar with charge 4/3. 
\end{itemize}

Because of the different initial states and production mechanisms, results from the Tevatron and LHC cannot be directly compared.  However, new physics models such as these can be used to make predictions at both machines, and therefore serve as a Rosetta stone of sorts to compare results.  As shown in Figure \ref{models_results}, most of the considered models still have available phase space that agree within the uncertainties of both the Tevatron and LHC charge asymmetry measurements. ATLAS and CMS measurements performed with the 8~TeV 20~$fb^{-1}$ data set are underway, and will lead to results with roughly equal-magnitude statistical and systematic uncertainties.  The 14-TeV run at the LHC will quickly produce results at twice the energy with larger datasets.  However, the asymmetry is predicted to be smaller than 14 TeV; therefore, the sensitivity of these measurements may be affected.  Although the anticipated results should be sufficient to severely constrain many new physics models, systematic uncertainties will remain a challenge in the measurement of the  charge asymmetry predicted by the SM at NLO QCD.


\subsection{\ttbar\ Spin Correlations}
\label{sec:spincorr}

The QCD hadronization scale ($\Lambda_{QCD} \sim $ 0.1 GeV) is much larger than the spin decorrelation scale ($\Lambda_{QCD}^2/m_t \sim$ 0.1 MeV). Therefore, any spin correlations in top quark pair production are reflected in angular correlations of the
decay products~\cite{Mahlon:1995zn,Stelzer:1995gc}.

\begin{figure}[htbp]
\begin{center}
\hspace*{-1.5cm}\includegraphics[width=1.2\textwidth]{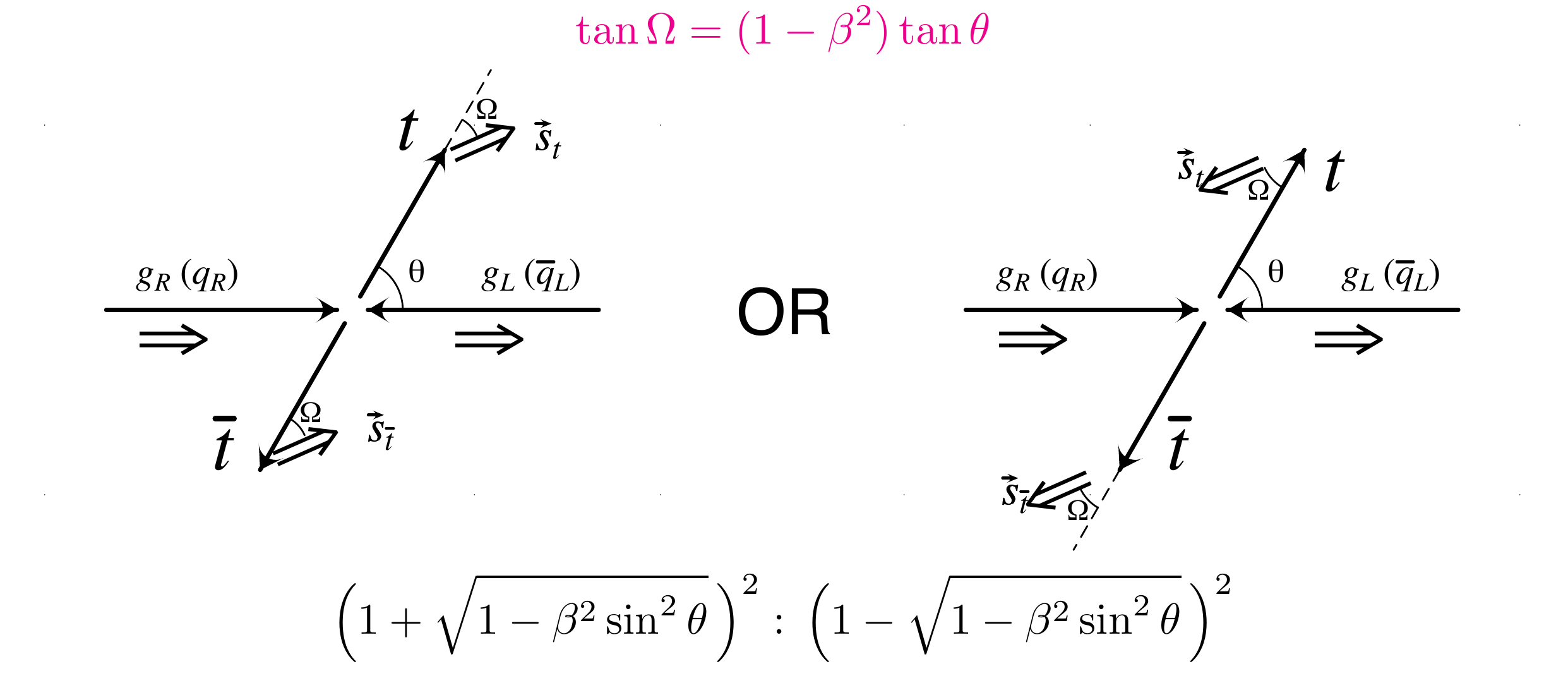}     
\end{center}
\vspace*{-0.5cm}
\caption{Spin correlations for the production of top quark pairs via unlike helicity gluon fusion or quark-antiquark annihilation.}
\label{fig:unlike}
\end{figure}

For top quark pair production via quark-antiquark annihilation or unlike-helicity gluon fusion, there exists a spin axis such that the top quarks are produced in only
the up-down or down-up configuration, namely parallel, given that the spin axes are back to back:
\begin{eqnarray}
q_L \bar{q}_R, ~q_L \bar{q}_R, ~g_L g_R, ~g_R g_L  \rightarrow  t_U \bar{t}_D + t_D \bar{t}_U.
\nonumber
\end{eqnarray}
No combinations $ t_U \bar{t}_U$ or $t_D \bar{t}_D$ are produced, see figure~\ref{fig:unlike}. This spin basis is known as the off-diagonal 
basis~\cite{Parke:1996pr,Mahlon:2010gw},
and the spin axis makes an angle $\Omega$ with respect to the top quark momentum direction in the zero mass frame (ZMF). In the ZMF, this angle is given by
\begin{eqnarray}
\tan \Omega  = (1-\beta^2) \tan \theta = \frac{1}{\gamma^2} \tan \theta
\nonumber
\end{eqnarray}
where $\beta$ and $\theta$ represent the speed and the scattering angle of the top quark, respectively.  Note that at threshold, $\Omega=\theta$ and the spin axis is aligned along the beam line, whereas at ultra-high energies, $\Omega=0$ and the spin axis is aligned along the direction of motion of the top quark.

\begin{figure}[htbp]
\begin{center}
\hspace*{-1.5cm}\includegraphics[width=1.2\textwidth]{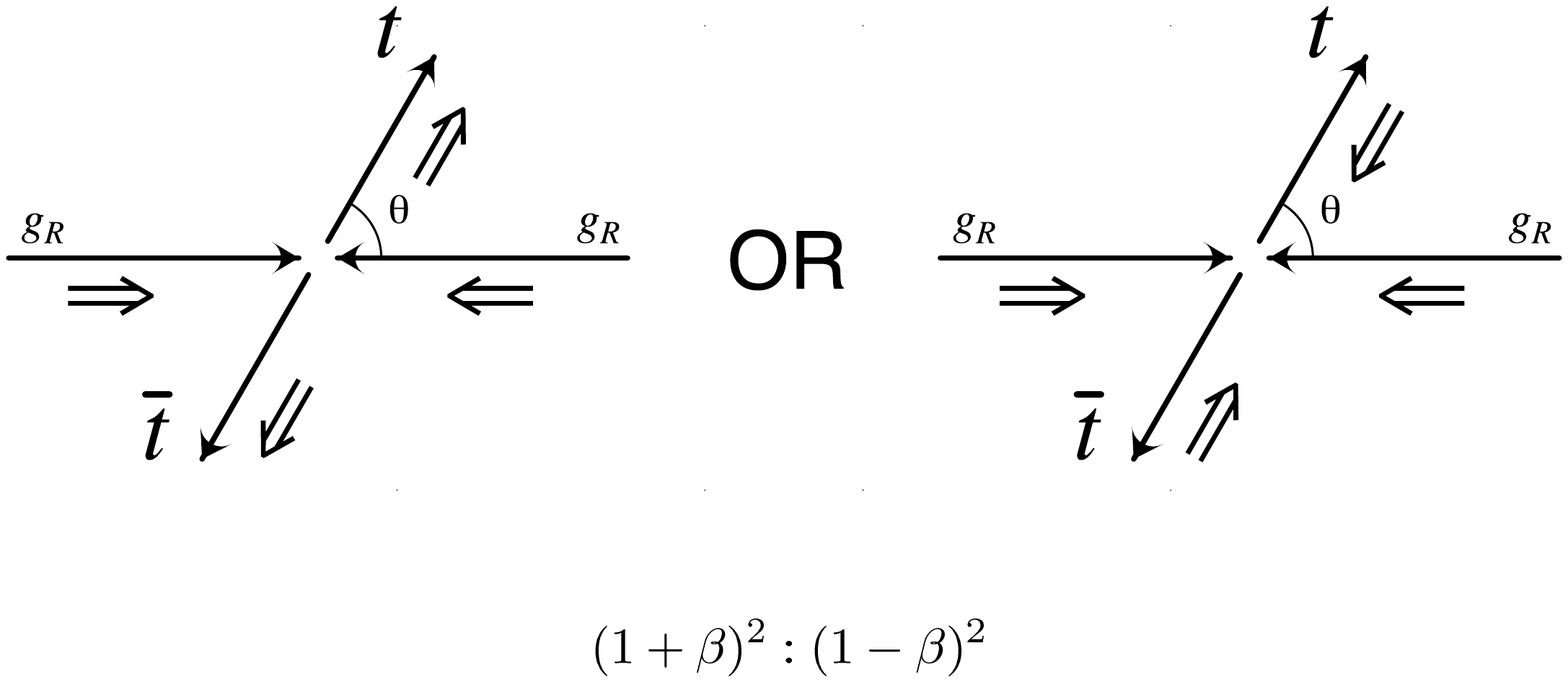}     
\end{center}
\vspace*{-3cm}
\caption{Spin correlations for the production of top quark pairs via like helicity gluon fusion.}
\label{fig:like}
\end{figure}

For top quark pair production via like helicity gluon fusion in the helicity basis, the top quarks are produced in only
the left-left or right-right configuration, that is, antiparallel:
\begin{eqnarray}
g_L g_L, ~g_R g_R  \rightarrow  t_L \bar{t}_L + t_R \bar{t}_R.
\nonumber
\end{eqnarray}
No combinations $ t_L \bar{t}_R$ or $t_R \bar{t}_L$ are produced in the like helicity gluon fusion process (Figure~\ref{fig:like}).

The dominant effect of the spin correlations is to correlate the angles of the decay products between the top quark and antitop quark,
that is, between $\chi_i$ and $\bar{\chi}_{\bar{i}}$. This correlation is given by
\begin{eqnarray}
\frac{1}{\sigma_T} \frac{d^2 \sigma}{d \cos \chi_i  ~d \cos \bar{\chi}_{\bar{i} }  }=  \frac{1}{4} (1+C_{t\bar{t}} ~\alpha_i \alpha_{\bar{i}} ~ \cos \chi_i \cos \bar{\chi}_{\bar{i}}) 
\nonumber
\end{eqnarray}
where the spin correlation coefficient, $C_{t\bar{t}}$ , is expected to have the following values in the SM:
\vspace{-0.5cm}
\begin{eqnarray}
C_{t\bar{t}} \equiv \frac{\sigma_{\uparrow \uparrow}+\sigma_{\downarrow \downarrow}-\sigma_{\uparrow \downarrow}-\sigma_{\downarrow \uparrow}}
{\sigma_{\uparrow \uparrow}+\sigma_{\downarrow \downarrow}+\sigma_{\uparrow \downarrow}+\sigma_{\downarrow \uparrow}}
=\left\{ \begin{array}{lll}
-0.456 & (-0.389) & {\rm Helicity ~at ~Tevatron}\\
+0.910 & (+0.806) & {\rm Beamline ~at ~Tevatron}\\
+0.918 & (+0.913) & {\rm Off-Diagonal ~at ~Tevatron}\\[0.2cm]
+0.305 & (+0.311) & {\rm Helicity ~at ~LHC(14 ~TeV)},
\end{array} \right. \nonumber
\end{eqnarray}
These values are at LO; NLO values are in parentheses~\cite{Bernreuther:2001rq}. At the LHC, the coefficient $ C_{t\bar{t}}$ in the off-diagonal and beamline bases is small, $<0.10$.

Both CDF and D0 have measurements of  these  spin correlation in top quark pair production.
CDF~\cite{Aaltonen:2010nz}, has measured the spin correlation coefficient in the helicity basis given by $C_{t\bar{t}}= 0.60 \pm 0.50 ~{\rm (stat.)} \pm 0.16 ~\rm{(syst.)}$, which is consistent with the QCD prediction, $C_{t\bar{t}} \approx 0.40$.
The D0~\cite{Abazov:2011qu, Abazov:2011gi}, measurement of this spin correlation coefficient is consistent with the SM, 
and  D0 has also applied a ME approach to this measurement. Combining these two measurements, D0 has obtained 3.1 $\sigma$ evidence for SM spin correlations in top quark pair production.


However, interference effects occur between the various spin components of the $t\bar{t}$ system, for example, between $t_L \bar{t}_L$ and $t_R \bar{t}_R$ for like-helicity gluon fusion, which leads to azimuthal correlations between the decay products:
\begin{eqnarray}
{1 \over \sigma_T}  {d\sigma \over d\Delta \phi}   = {1 \over 2} (1-D \cos \Delta \phi)
\nonumber
\end{eqnarray}
where in the ZM frame, the azimuthal correlations along the production axis are given by~\cite{Bernreuther:2010ny}
\begin{eqnarray}
D=\left\{ \begin{array}{ll}
+0.132  & {\rm Tevatron}\\
-0.353 & {\rm LHC(14~TeV)}.
\end{array} \right. 
\nonumber
\end{eqnarray}

\begin{figure}[ht]
\begin{minipage}[b]{0.45\linewidth}
\centering
\includegraphics[width=1.2\textwidth]{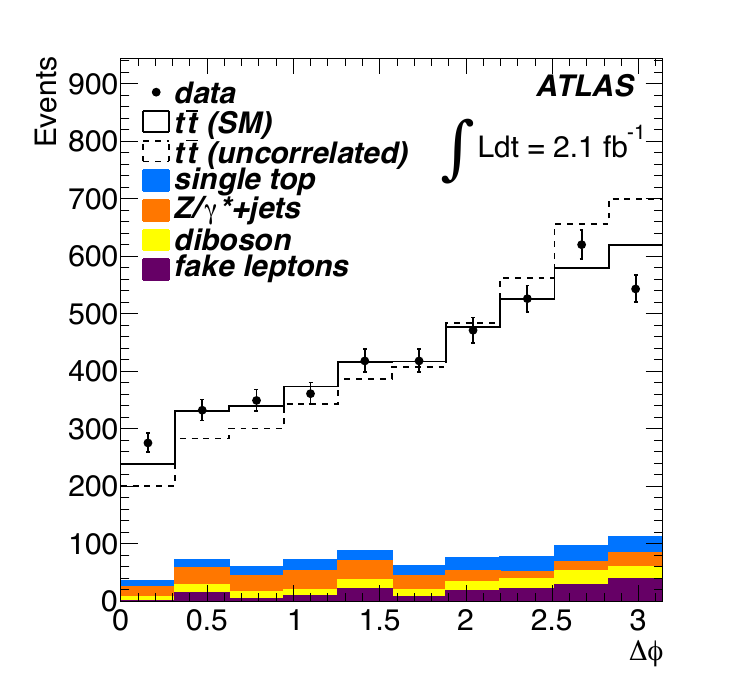}
\end{minipage}
\hspace{0.5cm}
\begin{minipage}[b]{0.45\linewidth}
\centering
\includegraphics[width=1.2\textwidth]{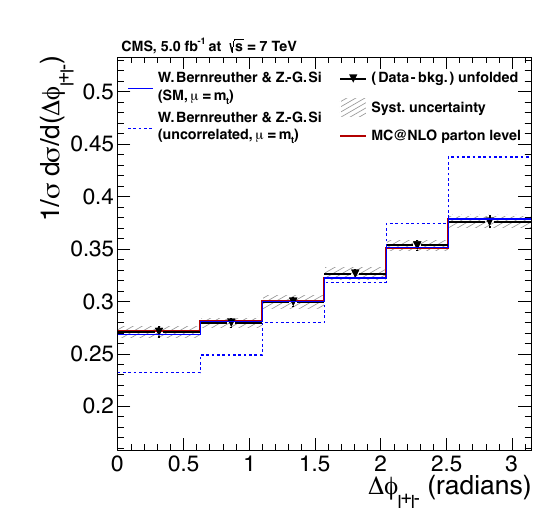}
\end{minipage}
\caption{The differential distribution of the Lab $\Delta \phi_{l^+ l^-}$ of the dilepton top quark pair events for ATLAS (left) and CMS (right).\label{fig:spincorr}}
\end{figure}

The azimuthal correlations about the beam axis in the Laboratory frame for the $\ell \ell$ events are discussed in Reference~\cite{Mahlon:2010gw}.  These azimuthal correlations are easier to observe
than the other angular correlations. In fact, they have been observed, at the 5 $\sigma$ level, by both the Atlas~\cite{ATLAS:2012ao} and CMS~\cite{Chatrchyan:2013wua} experiments at the LHC (Figure \ref{fig:spincorr}).

\subsection{\ttbar\ Production in association with Electroweak Boson}
\label{sec:ttV}

Although the top quark was discovered almost 20 years ago, we still know very little about its couplings to the electroweak bosons.
New physics connected with electroweak symmetry breaking can manifest itself as a deviation in these couplings from the SM prediction.
Models like technicolor or new strongly coupled Higgs bosons could alter the production of the top quark in association with vector bosons.
The measurement of $\ttbar \gamma$ cross section could also constrain the existence of excited top quark decaying into 
$t^* \to t \gamma$. Thus, measuring $t\bar{t}$ pairs in association with electroweak bosons is a key test of the validity
of the SM at the TeV scale. 
Moreover, note that the $\ttbar \gamma$ final state is an important control sample for the $\ttbar H$ 
process in which $H \to \gamma \gamma$. This process is an interesting production mode to measure the top quark-Higgs Yukawa coupling.
However, the cross sections of $t\bar{t}$ pairs in association with electroweak bosons 
are small, so to date only a limited number of such measurements have been
performed at colliders.

CDF has measured the ratio of the $\ttbar \gamma$ to \ttbar\ production cross sections. This ratio allows for cancellation of some of the
systematic effects and thus is more sensitive than the measurement of the $\ttbar \gamma$ cross section alone~\cite{Aaltonen:2011sp}.
The analysis was performed in the $\ell$+jets channel using 6~fb$^{-1}$ of data. The $\ttbar \gamma$ process was selected with 
the same kinematic cuts than the \ttbar\ final state but with a requirement for an isolated photon with transverse energy of 
more than 10~GeV. 
The largest background comes from jet misidentified as photon and is evaluated in data using control samples. 
CDF finally measures $\sigma_{\ttbar \gamma} = 0.18 \pm 0.07 {\rm (stat)} \pm 0.04 {\rm (syst)} \pm 0.01 {\rm (lumi)} $~pb 
corresponding to the first evidence (3.0 $\sigma$) for this process. The result is agrees with the SM prediction of $0.071 \pm 0.011$~pb. 

\begin{center}
\begin{figure}[htbp]
\hspace*{-3cm}\includegraphics[width=1.0\textwidth]{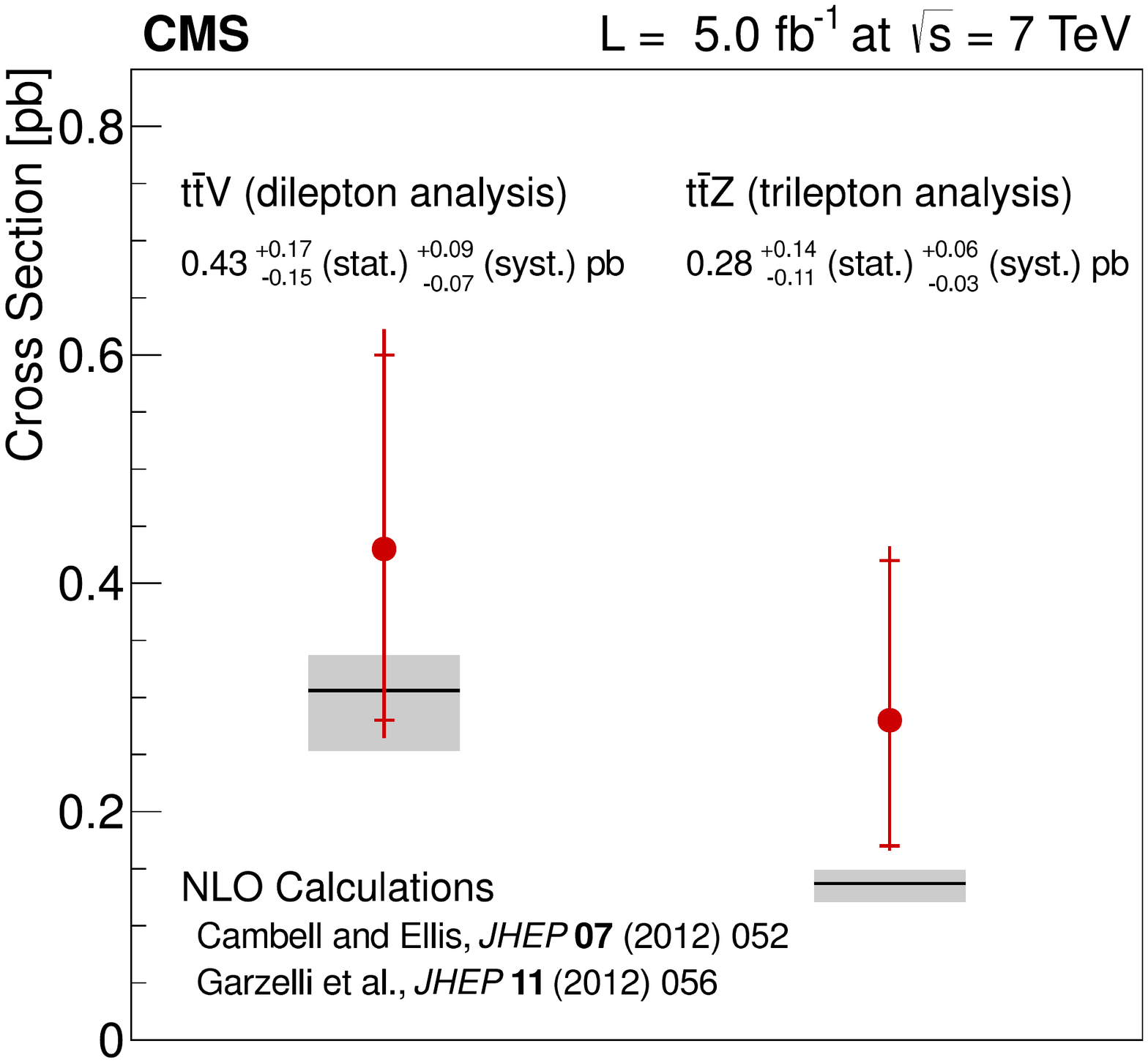}
\caption{Summary of the CMS $\ttbar Z$ and $\ttbar V$ cross-section measurements. The Standard Model predictions (gray boxes) are shown for comparison~\cite{Chatrchyan:2013qca}.  }
\label{fig:cmsttV}
\end{figure}
\end{center}

CMS, using 5~fb$^{-1}$ of data, published the first measurement of vector boson production associated with a \ttbar\ pair ~\cite{Chatrchyan:2013qca}.
This analysis was conducted with $\ell \ell \ell$ and same-sign $\ell \ell$ signatures. The $\ell \ell \ell$ final state could originate
from $\ttbar Z$ in the $\ell$+jets channel, whereas $Z \to \ell \ell$
(where $\ell = e$ or $\mu$). In this channel, two same-flavor, opposite-charge electrons or muons are selected with an
invariant mass close to the $Z$ mass. The third lepton is required to have a lower (> 10~GeV)$p_T$ than that of the other two
(> 20~GeV). The main background comes from Drell-Yan process with an additional 
lepton reconstructed from hadronization products and WZ events.
The same-sign dilepton analysis searches for $\ttbar V$ events from $\ttbar W$ in the $\ell$+jets channel. 
$\ttbar V$ events can also originate from $\ttbar Z$ with a decay chain like in the $\ell \ell \ell$ channel described above. 
The same-sign final state benefits from little background from SM processes 
($WZ$, $ZZ$, $Z\gamma$, $W\gamma$, $WW$). 
The dominant background in this channel comes from nonprompt leptons or misreconstruction effects.
Both the $\ttbar Z$ and $\ttbar V$ cross section measurements are currently limited by statistics and are compatible with the
SM predictions. 
Figure~\ref{fig:cmsttV} shows the results. The $\ttbar Z$ ($\ttbar V$) signal has been established 
with a significance of 3.3$\sigma$ (3.0$\sigma$).

\section{Conclusion}
\label{sec:conclusion}

The accuracy of the measurements of properties of the top quark has improved steadily
for the past two decades, as the 
number of top quarks available for study has increased from the few tens used for discovery to tens of thousands now.
The mass of the top quark has been determined to an impressive accuracy of better than 1\%.  The accuracy of the measurements of many other properties is approaching the few-percent level as well. Despite a few tantalizing hints, all the properties of the top quark
remain consistent with expectations from the SM. 
With the approaching increase in the energy and
luminosity of the LHC, significantly larger samples of top quarks will become available in a few years,
and studies of the properties of top quarks will continue to improve.
Whether the top quark will play an important role
in any discovery of new physics remains to be determined.  However, given its unique particles properties, the top quark will remain an important focus in studying SM particle physics and in searches for physics beyond it.


\end{document}